\documentclass[aps,pre,reprint]{revtex4-1}
\usepackage[T1]{fontenc}
\usepackage{amsmath}
\usepackage[latin1]{inputenc}
\usepackage{graphicx}

\usepackage{epsfig}

\usepackage{color}

\begin{document}
\title{Sedimentation stacking diagram of binary colloidal mixtures\\ and
bulk phases in the plane of chemical potentials}

\author{Daniel de las Heras}
\affiliation{Theoretische Physik II, Physikalisches Institut,
 Universit{\"a}t Bayreuth, D-95440 Bayreuth, Germany}

\author{Matthias Schmidt}
\email{Matthias.Schmidt@uni-bayreuth.de}
\affiliation{Theoretische Physik II, Physikalisches Institut,
 Universit{\"a}t Bayreuth, D-95440 Bayreuth, Germany}

\date{\today}
\begin{abstract}
We give a full account of a recently proposed theory that explicitly
relates the bulk phase diagram of a binary colloidal mixture to its
phase stacking phenomenology under gravity [{\it Soft Matter {\bf 9},
    8636 (2013)}]. As we demonstrate, the full set of possible phase
stacking sequences in sedimentation-diffusion equilibrium originates
from straight lines (sedimentation paths) in the chemical potential
representation of the bulk phase diagram. From the analysis of various
standard topologies of bulk phase diagrams, we conclude that the
corresponding sedimentation stacking diagrams can be very rich, even
more so when finite sample height is taken into account. We apply the
theory to obtain the stacking diagram of a mixture of nonadsorbing polymers
and colloids. We also present a catalog of generic phase diagrams in the plane of chemical
potentials in order to facilitate the practical application of our
concept, which also generalizes to multi-component mixtures.
\end{abstract}

\maketitle

\section{Introduction \label{introduction}}

Binary colloidal mixtures are fascinating objects of study for a
variety of reasons, including their role as model systems for more
complex industrial products, but also for their suitability to
investigate fundamental phenomena in condensed matter, such as the
occurrence of complex phase behaviour~\cite{tuinier2003depletion,leunissen2005ionic,C1SM06948A,varrato2012arrested,vermolen}. In particular using
(well-controlled) particle shapes of one or both colloidal components
opens up a vast phase phenomenology of collective ordering phenomena
both in bulk and at interfaces.

Carrying out systematic studies of binary mixtures is often
significantly more challenging than studying corresponding
one-component systems. As the presence of gravitational effects in the
lab is often unavoidable, these are commonly used as a tool to study
the collective behaviour of colloidal mixtures~\cite{C1SM06535A}. The
theoretical modelling of sedimentation via height-dependent density
profiles~\cite{biben1993density} that are caused by the external
gravitational potential is in principle straightforward. Nevertheless,
drawing systematic conclusions from observations of the phenomena
under gravity about the (often unknown) bulk phase diagram constitutes
a highly demanding task.  Furthermore, besides being a primary
experimental tool, the analysis of sedimentation also reveals
genuinely new phenomena, such as e.g.\ the emergence of floating
phases~\cite{floating}, and zone formation due to
  interplay between sedimentation and phase ordering
  \cite{0295-5075-89-3-38006}. For a recent review on sedimentation
see~\cite{PiazzaReview}.

Parola and coworkers recently developed a microscopic theory~\cite{Piazza1}, based on
Statistical Mechanics, of the buoancy effects that are due to the
``granular'' character of a suspending colloidal dispersion. Here the
``solute'' colloid is different in size and its properties from the
``bath'' colloids. As Parola et al. demonstrate very convincingly, a
modified, generalized Archimedes principle applies in this case. On
its basis, the authors rationalize the sedimentation behavior of a
dilute component in a dense colloidal background of a primary
component, as is a very relevant case, cf. the body of experimental
work by Piazza and coworkers \cite{Piazza1,Piazza2}.

Our recently proposed approach~\cite{stackingdiagram} complements the work of Parola et
al. in that we do not impose the restriction of infinite dilution of
the solute colloid, which enables us address phase transitions in the
full binary mixture (not just in the background
component). Furthermore, our approach is generic in the sense that the
bulk phase diagram is considered as input to the theory.  We place
particular emphasis on the chemical potential representation of the
bulk phase diagram, which we find to be particularly well suited for
the analysis of sedimentation in dense systems.  The theory applies
generically, whether the bulk phase diagram originates from a
microscopic treatment, such as the approach by Parola and coworkers~\cite{Piazza1,Piazza2},
or classical density functional theory, or approaches such as the free
volume theory \cite{0295-5075-66-1-125} (which we employ in one of the
cases studied below).

The present contribution serves several purposes.  First, we extend
our previous short account~\cite{stackingdiagram}: we investigate the role of an
inflection point in the bulk binodal; we expand the description of an
alternative thermodynamic variable set ($r_{min}$,$\alpha$ plane), and
show results for both "realistic" and "complex" stacking diagram.

Second, we show the stacking diagram for a mixture of polymers and
platelets with both infinite and finite sedimentation paths. Here we
show the stacking diagram for finite paths in several planes in order
to demonstrate the equivalence of the chemical potential
representation with other, more common representations for the bulk phase diagram.

Third, we show schematically for a variety of relevant topologies the
correspondence between the pressure-composition and the chemical
potential representation of the bulk phase diagrams.

\section{Theory \label{theory}}
\subsection{The sedimentation path}\label{sedpath}

Consider a binary colloidal mixture in a given solvent and in presence of gravity. In sedimentation-diffusion equilibrium, we can define a height-dependent local chemical potential~\cite{schmidt04aog,floating,stackingdiagram} for each species $i={1,2}$:
\begin{eqnarray}
\psi_1(z)&=&\mu_1^b-m_1gz,\nonumber\\
\psi_2(z)&=&\mu_2^b-m_2gz,\label{eq:local}
\end{eqnarray}
where $\mu_i^b$ is the bulk chemical potential of species $i$, $m_i$ is its buoyant mass, $z$ is the vertical coordinate, and $g$ is the acceleration of gravity. Eliminating the spatial dependency of the local chemical potentials in Eq. (\ref{eq:local}) results in
\begin{equation}
\psi_2(\psi_1)=a+s\psi_1,\label{eq:sedpath}
\end{equation}
where both $a$ and $s$ are constants given by
\begin{eqnarray}
a&=&\mu_2^b-s\mu_1^b,\nonumber\\
s&=&m_2/m_1.\label{eq:constants}
\end{eqnarray}
Eq. (\ref{eq:sedpath}) represents a line segment, the sedimentation path, in the plane of local chemical potentials. The sedimentation path describes the variation of local chemical potentials along the sedimented colloidal mixture. 

Next, we introduce a local density approximation (LDA)~\cite{stackingdiagram} assuming that locally, i.e. at any $z$, the state of the system is the same as an equilibrium bulk state with chemical potentials $\mu_i$ that equal the local chemical potentials, i.e.:
\begin{equation}
\mu_i=\psi_i(z).\label{LDA}
\end{equation}
This approximation is justified if the relevant correlation lengths in the system are small compared to the gravitational lengths of the colloids $\xi_i=k_BT/(m_ig)$, where $k_B$ is the Boltzmann constant and $T$ is the absolute temperature. In colloidal systems $\xi$ is typically of the order of millimeters or centimeters. Hence, the LDA can often be an accurate approximation for the system under consideration. 

\begin{figure}
\epsfig{file=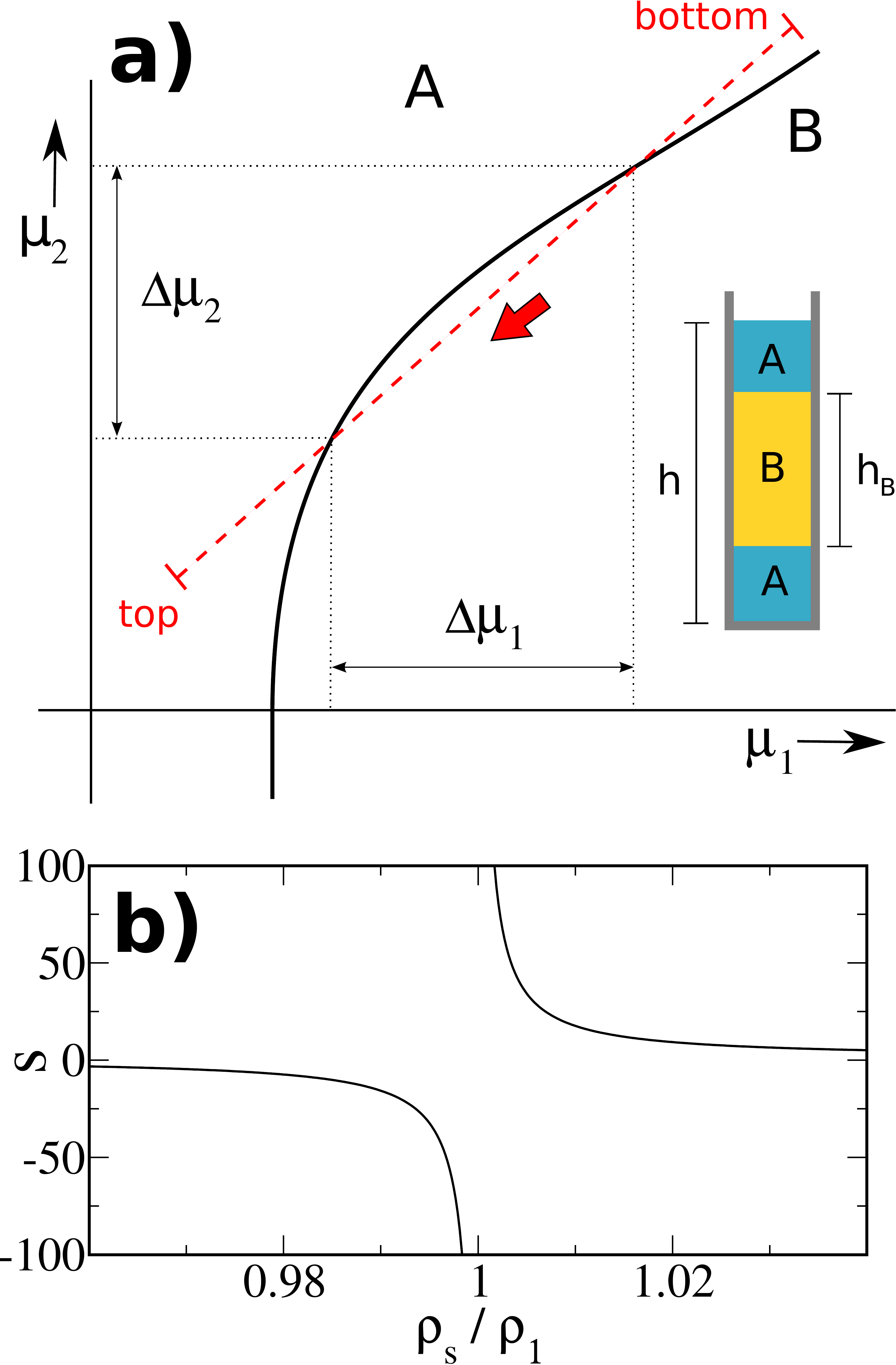,width=0.8\columnwidth,clip=}
\caption{a) Schematic bulk phase diagram of a colloidal binary mixture in the plane of chemical potentials $\mu_1$,$\mu_2$. The black-solid line is a binodal where phases A and B coexist. The red-dashed line is a sedimentation path, the direction of which (from bottom to top) is specified by an arrow. The inset represents the stacking sequence corresponding to the sedimentation path: ABA. The differences in chemical potentials $\Delta\mu_i$ are proportional to the thickness of the middle stack $h_B$. b) Slope of the sedimentation path $s=m_2/m_1$ as a function of the solvent density scaled with the mass density of species $1$, $\rho_s/\rho_1$. The curve is qualitatively the same independently of the mass densities $\rho_i$ and particles volumes $v_i$ provided that $\rho_1\neq\rho_2$. If $\rho_1=\rho_2$ the slope of the sedimentation path is constant and equal to the ratio between the particle volumes.}
\label{fig1}
\end{figure}

The LDA allow us to directly relate the sedimentation path, Eq. (\ref{eq:sedpath}), and the stacking sequence observed in the sample because an interface in the sample corresponds to a crossing between the path and a binodal in the chemical potential representation of the bulk phase diagram. In order to illustrate this point, we show in Fig. \ref{fig1}a) a schematic bulk phase diagram of a binary mixture in the plane of both chemical potentials. In bulk there are two stable phases, A and B, that coexist along the binodal. The sedimentation path starts (bottom) in the region where A is stable. Then it crosses the binodal entering the region of stability of B. Finally it crosses again the binodal and ends (top) in the region of stability of A. The corresponding stacking sequence is ABA, i.e., bottom A, middle B, and top A. In what follows we label the sequence of stacks from bottom to top of the sample. Paths that cross a binodal twice, such as the one shown in Fig. \ref{fig1}a), were recently used in order to rationalize the experimental observations in a mixture of gibbsite platelets and silica spheres~\cite{floating}. 

The (macroscopic) distance between two heights in the sample can be directly transformed into the difference of chemical potentials in the bulk phase diagram via Eq. (\ref{eq:local}). For example, the difference in chemical potentials, $\Delta\mu_i$, between the crossing points of the binodal and the path in Fig. \ref{fig1}a) is proportional to $h_B$, the thickness of the B stack in the sample, $\Delta\mu_i=-m_igh_B$. Hence, the difference in chemical potentials between two heights in the sample is an experimentally accessible quantity.

A sedimentation path is fully specified by its slope, length, direction and position in the chemical representation of the bulk phase diagram. Each of these variables is related to physical parameters of the system: 

\begin{itemize}
\item The slope, $s$, of a sedimentation path is given by the ratio of the buoyant masses, cf. Eq. (\ref{eq:constants}), or by the inverse ratio of the gravitational lengths $s=\xi_1/\xi_2$. The buoyant mass is $m_i=(\rho_i-\rho_s)v_i$, with $\rho_s$ the solvent density, and $\rho_i$ and $v_i$ the mass density and particle volume of species $i$, respectively. Hence the slope of the path can be written as:
\begin{equation}
s=\frac{(\rho_2-\rho_s)v_2}{(\rho_1-\rho_s)v_1}.
\end{equation} 
Therefore, one can experimentally control $s$  by e.g. varying the density of the solvent, provided that $\rho_1\neq\rho_2$, see Fig. \ref{fig1}b). Alternatively, one could vary $s$ by varying the buoyant mass of one species by, e.g., designing colloids with inner cores made of a different material \cite{doi:10.1021/la0101548}.

\item The length of the path in the $\mu_1$,$\mu_2$ plane is fixed by the buoyant masses and the height of the container, $h$. The difference in local chemical potentials from top to bottom of the sample is, cf. Eq. (\ref{eq:local}), $\Delta\mu_i^T=-m_igh$.

\item The direction of a sedimentation path is determined by the signs of the buoyant masses. If $m_i$ is positive (negative) then $\mu_i(z)$ decreases (increases) from bottom to top of the sample. In the example of Fig. \ref{fig1}a) both, $m_1$ and $m_2$, are positive. Again by changing the density of the solvent it would be possible to change the sign of the buoyant mass.

\item The location of the path in the plane of chemical potentials is given by the bulk chemical potentials of the sample in absence of gravity. Hence, the location is determined by the overall composition and concentration of the mixture. 

\end{itemize}

\begin{figure}
\epsfig{file=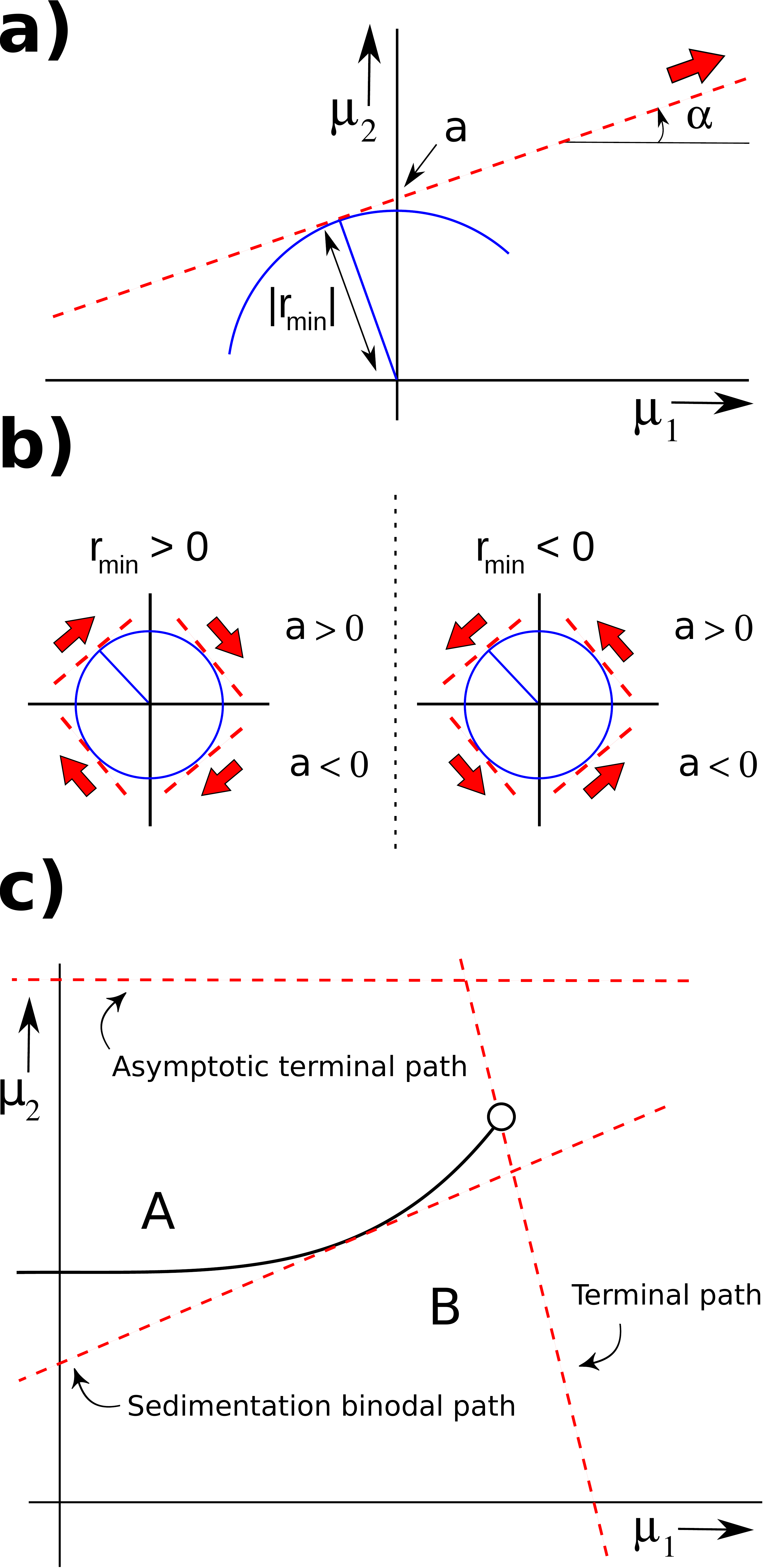,width=2.5in,clip=}
\caption{(a) A sedimentation path (red-dashed line) in the plane of chemical potentials $\mu_1$,$\mu_2$; the arrow indicates its direction. $\alpha$ is the angle between the path and the $\mu_1$-axis. $a$ is the $y$-value of the path at the intersection with the $\mu_2$-axis. $\lvert r_{min}\rvert$ is minimum distance between the path and the origin. (b) Sing convention for $r_{min}$: positive (left) for a clockwise path and negative (right) for an anticlockwise path. The sign convention allow us to distinguish paths with the same $\alpha$ but opposite $a$. (c) Schematic bulk phase diagram in the plane of chemical potentials $\mu_1$,$\mu_2$. Two phases A and B coexist along the binodal curve (black-solid line). The binodal starts at a phase transition of the species $2$ (horizontal asymptote) and ends at a critical point (empty circle). Three sedimentation paths are shown (red-dashed lines). They all correspond to boundaries between two stacking sequences in the stacking diagram. An infinitesimal change of the path changes the stacking sequence.}
\label{fig2}
\end{figure}

Each sedimentation path is associated with a corresponding phase stacking sequence. The set all possible stacking sequences for a given bulk phase diagram forms the stacking diagram. The phase space of the stacking diagram is that of the sedimentation path, i.e., slope, length, position and direction of the path. Hence, the stacking diagram is a multidimensional object. However, in a standard sedimentation experiment of colloidal mixtures, the height of the sample is $h\sim 1 \text{ cm}$. This implies a length of the sedimentation path of several $k_BT$ \cite{floating} that typically covers the whole bulk phase diagram of the mixture. This allow us to simplify the representation of the stacking diagram by neglecting effects due to the finite height of the sample (we will mention the effect of having a finite sample latter). Under this approximation a sedimentation path is a straight line (instead of a line segment) that can be described using the parameters $a$ and $s$, see Eq. (\ref{eq:constants}) and Fig. \ref{fig2}a). Both $a$ and $s$, together with the direction of the path, are our choice for the parameter space of the stacking diagram. Alternatively, we also use $\alpha$ and $r_{\text{min}}$ as a different parameter space, see Fig. \ref{fig2}a). $\alpha$ is the angle between the path and the $\mu_1$ axis. $\lvert r_{\text{min}}\rvert$ is the minimum radius of a circle centered at the origin of the chemical potentials and tangent to the path. In order to distinguish between two paths with the same value of $\alpha$ but opposite values of $a$ we ascribe a sign to $r_{\text{min}}$, which we choose to be positive (negative) for a clockwise (anticlockwise) path, see Fig. \ref{fig2}b).

\subsection{The stacking diagram}

In order to construct the stacking diagram we need to find the boundaries between distinct stacking sequences in the $a-s$ parameter space. There are three types of boundaries, which we refer to as sedimentation binodals, terminal lines and asymptotic terminal lines, as described in the following.\\ 

{\bf (i) Sedimentation binodal.} A sedimentation path tangent to a binodal in the bulk phase diagram is a special path. Any infinitesimal variation of the values $a$ or $s$ of the path changes the corresponding stacking sequence, see Fig. \ref{fig2}c). The set of all paths (represented by $a$ and $s$) that are tangent to a bulk binodal forms the sedimentation binodal. Let $\mu_{2,\rm AB}(\mu_1)$ be the parameterization as a function of $\mu_1$ of the chemical potential of species $2$ at bulk AB phase coexistence. The sedimentation paths that are tangent to the AB binodal then are those that satisfy
\begin{equation}
a_{\rm AB}(s)=\mu_{2,\rm AB}(\mu_1)-\mu_1 s,\label{eq:legendre}
\end{equation}
where the slope, $s$, of the path is given by
\begin{equation}
s=\frac{d\mu_{2,\rm AB}}{d\mu_1}.
\end{equation}
Eq. (\ref{eq:legendre}) constitutes the Legendre transform of the bulk binodal in the chemical potential representation.

{\bf (ii) Terminal line}. A sedimentation path that crosses a critical point constitutes a further special case. An example of such a path is depicted in Fig. \ref{fig2}c). The stacking sequence changes by infinitesimally varying the intercept of the path, $a$. The sequence is, however, robust against any changes in the value of $s$. In fact, any path that crosses an end point of a binodal (which can be a critical point, triple point, tricritical point, critical end point etc.) is special in the same sense and corresponds to a boundary between two different sequences in the stacking diagram. Let $\mu_{i,\rm end}$ be the chemical potential of species $i$ at an ending point of a binodal. Then the relation
\begin{equation}
a_{\rm end}(s)=\mu_{2,\rm end}-\mu_{1,\rm end}s,
\label{EQterminalLine}
\end{equation}
describes all the sedimentation paths that cross the end point.  Equation \eqref{EQterminalLine} describes a line in the $a-s$ plane, which we refer to as the terminal line. 

{\bf (ii) Asymptotic terminal line}. A bulk binodal does not terminate at finite chemical potentials, if it is connected to a phase transition of one of the pure components of the system. In this case the binodal has a horizontal or vertical asymptote in the plane of chemical potentials. Furthermore, a binodal that represents a demixing region at high chemical potentials does not terminate at finite chemical potentials either. In this case, the binodal tends also to an asymptote with a well defined slope, which can be rationalized as follows. The slope of a binodal representing the coexistence between phases A and B is given by:
\begin{equation}
\frac{d\mu_{2,{\rm AB}}}{d\mu_1}=-\frac{\Delta\rho_{1,{\rm AB}}}{\Delta\rho_{2,{\rm AB}}},
\end{equation}
where $\Delta\rho_{i,{\rm AB}}$ is the density jump of species $i$ at AB coexistence. At very high chemical potentials both species will reach the close packing densities and the slope of the binodal will be constant.

Hence all binodals that do not terminate at finite values of the chemical potentials tend to an asymptote with a well define slope,
\begin{equation}
\frac{\mu_{2,{\rm AB}}}{\mu_{1,{\rm AB}}}\rightarrow s_\infty.
\end{equation} 
A sedimentation path that is parallel to an asymptote of a binodal constitutes a boundary between two different stacking sequences in the stacking diagram, because a change in the slope of the path modifies the sedimentation state.  An example of such a path is shown in Fig. \ref{fig2}c). In the stacking diagram, all paths with $s(a)=s_\infty=\text{const}$, describe a line, which we refer to as the asymptotic terminal line.

\section{Results}

We apply our theory to obtain the stacking diagrams that corresponds to different model bulk phase diagrams. The bulk phase diagrams that we use do not stem from a microscopic treatment of a model Hamiltonian. They are rather intended to provide relevant examples of the stacking diagrams. Technically, we model the bulk binodals in the plane of chemical potentials using B\'ezier curves.

\subsection{Binodal ending at two critical points}
\begin{figure}
\epsfig{file=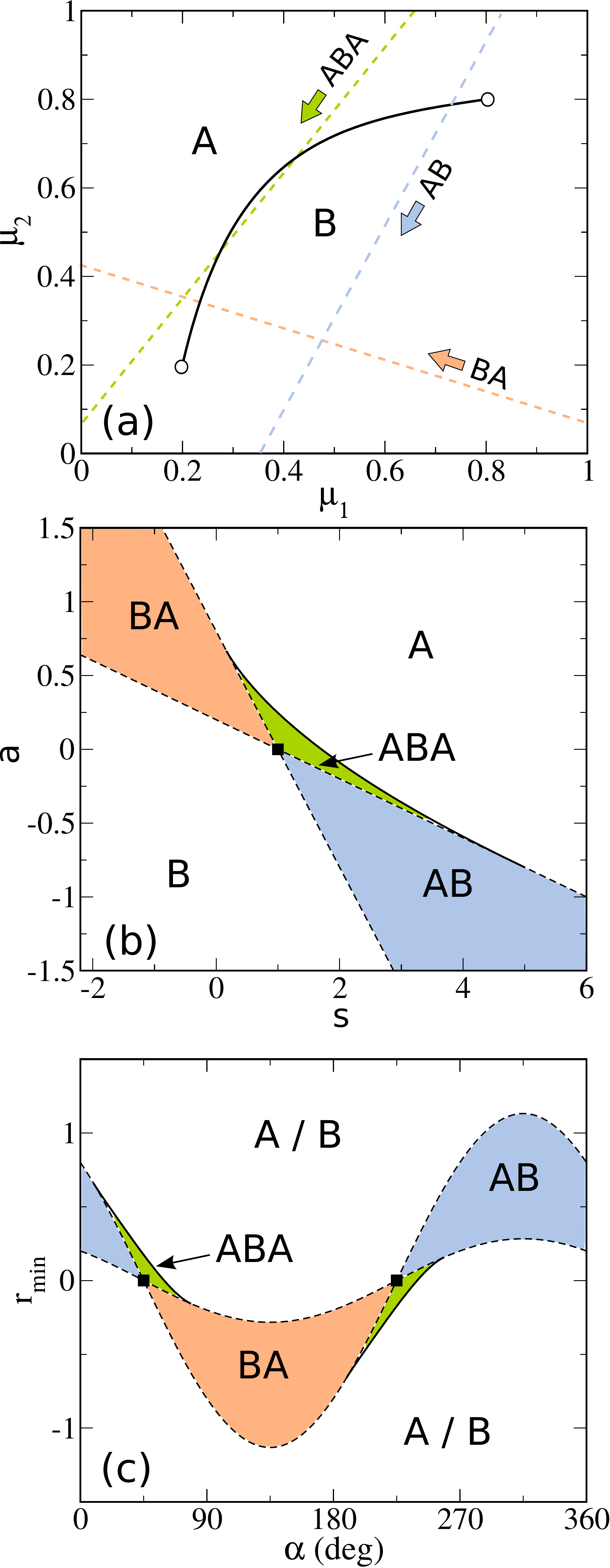,width=2.5in,clip=}
\caption{(a) Schematic bulk phase diagram of a binary mixture in the plane of chemical potentials $\mu_1$, $\mu_2$. Two phases, A and B, coexist along the binodal line (black-solid line). The empty circles are critical points. The dashed lines are representative sedimentation paths (the arrows give the directions from bottom to top of the sample and the labels indicates the stacking sequence). Chemical potentials are given in arbitrary units. (b) Phase stacking diagram in the plane of slope $s$ and intercept of the sedimentation path. Here we set $m_1>0$. (c) Phase stacking diagram in the $\alpha$, $r_{\text{min}}$ plane. In (b) and (c) the different stacking regions are coloured and the labelled (from bottom to top) by their respective stacking sequences. Sedimentation binodals are represented by black-solid lines and terminal lines are depicted as dashed lines. The black squares in (b) and (c) indicate the path that crosses both critical points in (a).}
\label{fig3}
\end{figure}

In Fig.~\ref{fig3}a) we show a  bulk phase diagram with the simple topology of a single binodal that ends at two critical points. Phases A and B coexist along the binodal. The stacking diagram, (b) and (c), consists on a sedimentation binodal (formed by those paths tangent to the bulk binodal) and two terminal lines (formed by those paths that cross the critical points). The terminal lines and the sedimentation binodal divide the stacking diagram in different regions. Each region correspond to a qualitatively different stacking sequence. The stacking sequence of each region can be found as follows: (i) select a point in the stacking diagram that lies in the region of interest, (ii) plot the corresponding sedimentation path in the bulk phase diagram, and (iii) obtain the stacking sequence by finding the crossing points between the path and the binodal (see a selection of different paths in Fig.~\ref{fig3}a). Two paths with the same slope and same intercept, but opposite directions, give rise to stacking sequences with reversed order (e.g., AB and BA). In order to avoid this ambiguity when representing the stacking diagram in the plane of the slope and intercept, we show only those paths with $m_1>0$ in Fig.~\ref{fig3}a). To obtain the stacking diagram in the case $m_1<0$, one only needs to reverse the staking sequences. The alternative $\alpha$,$r_{\text{min}}$ representation of the stacking diagram, Fig.~\ref{fig3}b), includes all possible cases (not only $m_1<0$), because we allow $\alpha\in[0,2\pi]$. Each bulk binodal generates two sedimentation binodals in the $\alpha$,$r_{\text{min}}$ representation of the stacking diagram, because a path tangent to a binodal has two directions, given by the pair $\alpha$ and $\alpha+\pi$. The points in the stacking diagrams where two terminal lines intersect are paths that cross two ending points of a binodal in the bulk phase diagram.

This basic example shows already the potential richness of the stacking diagram. Although there are only two stable phases in bulk, the stacking diagram contains up to five different stacking sequences. One of them is the sequence ABA that originates whenever a path crosses the bulk binodal twice. A similar phenomenon was theoretically predicted and experimentally found in mixtures of gibbsite platelets and silica spheres~\cite{floating}. Note that two consecutive phases in the stacking sequence must display coexistence in the bulk phase diagram. However, as we learn from this example, two non-consecutive phases may or may not coexist in bulk. Moreover, in the example the bottom and top phases of the ABA sequence are the same, which constitutes a reentrant phenomenon.

\subsection{The role of an inflection point}
\begin{figure}
\epsfig{file=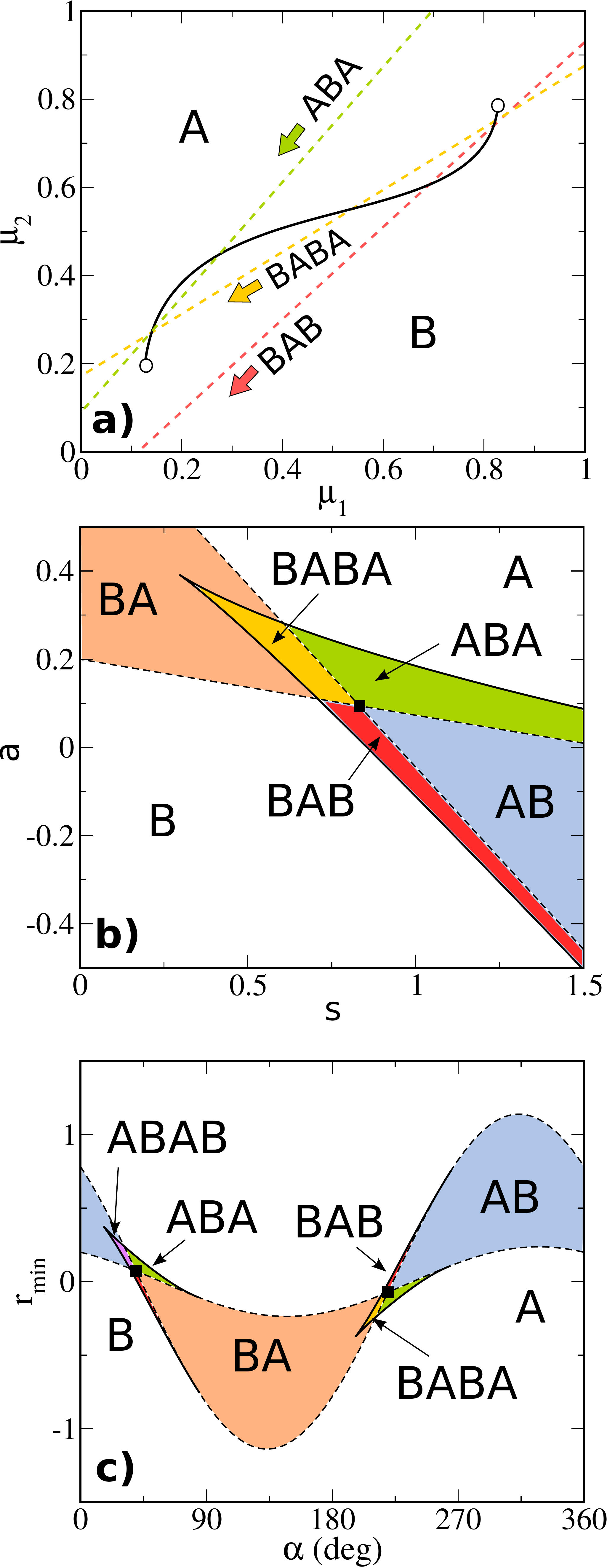,width=2.5in,clip=}
\caption{(a) Schematic bulk phase diagram of a mixture in the plane of chemical potentials, $\mu_1-\mu2$. Stacking diagram of the mixture in the $s$,$a$ (b) and $\alpha$,$r_{\text{min}}$ (c) planes. See caption of Fig.~\ref{fig3} for a full description.} 
\label{fig4}
\end{figure}

The curvature of the bulk binodal in the chemical potential representation plays an important role in determining the stacking diagram. Besides affecting the shape of the corresponding sedimentation binodal, the curvature can also affect the topology of the stacking diagram. In Fig.~\ref{fig4}a) we show a schematic bulk phase diagram in which a binodal ends at two critical points, as in the previous example. However, a change in the curvature of the binodal introduces an inflection point. An inflection point in a binodal has been recently predicted in mixtures of colloidal platelets  and spheres~\cite{floating} and it may be a common feature of binodals that connect two transitions of the pure components of a mixture~\cite{stackingdiagram}. The presence of an inflection point generates additional stacking sequences, because it increases the number of possible intersections between a sedimentation path and the binodal that contains the inflection point. The path BABA in Fig.~\ref{fig4}a) which crosses the binodal three times is an example. The stacking diagram of this mixture is depicted in panels (b) and (c) of Fig.~\ref{fig4}. As in the previous example, the stacking diagram possesses one sedimentation binodal and two terminal lines. There are in total eight possible stacking sequences. Due to the presence of the inflection point the sedimentation binodal is multivalued; there are sedimentation paths tangent to the binodal that share the same slope, $s$, but have different intercept, $a$. The sedimentation binodal has also a kink, which corresponds to the path tangent to the binodal at the inflection point. 

\subsection{Transitions in the pure subsystems of the mixture}
\begin{figure*}
\epsfig{file=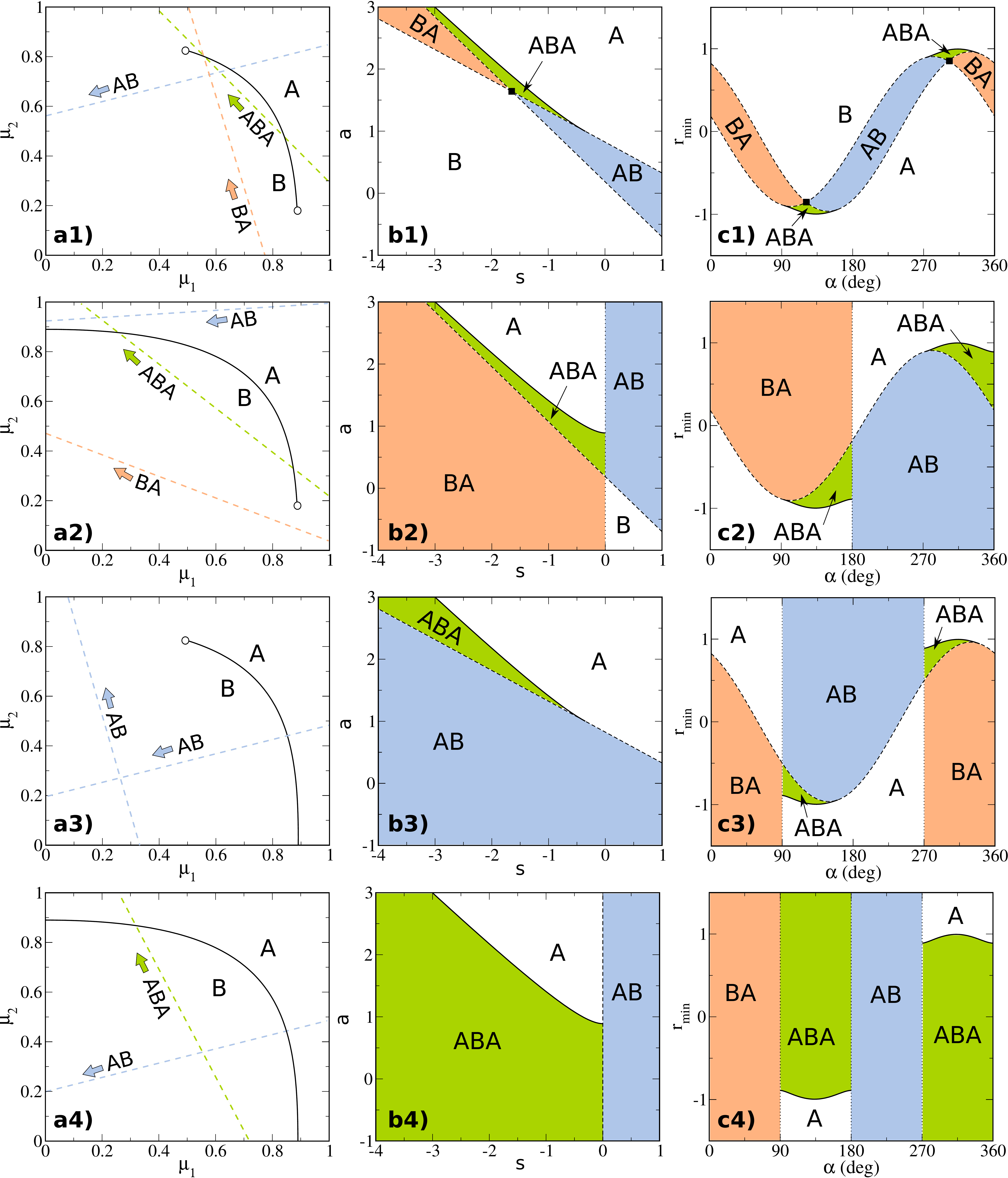,width=6.in,clip=}
\caption{Schematic bulk phase diagrams of a mixture in the plane of chemical potentials $\mu_1$,$\mu_2$ (first column) and their corresponding stacking diagrams in the plane of slope, $s$, and intercept, $a$, of the sedimentation path (second row) or in the $\alpha$,$r_{min}$ plane (third column). Chemical potentials are given in arbitrary units. In the second column we use $m_1>0$. There are two phases A and B that coexist along a binodal that: ends at two critical points (a1); ends at one critical point and at a phase transition in the pure system of species $2$ (a2) or species $1$ (a3); connects two phase transitions of the pure components of the mixture (a4). The solid black lines are the bulk (first column) and sedimentation (second and third column) binodals. Empty circles are critical points. Dashed (dotted) black lines are terminal lines (asymptotic terminal lines). The regions in the stacking diagrams have been coloured and labeled according to their stacking sequences. Selected sedimentation paths have been represented in the bulk phase diagram with an arrow indicating their direction and have been labelled according to their stacking sequence.}
\label{fig5}
\end{figure*}
We next consider that at least one of the pure components of the mixture undergoes a phase transition in bulk. In Fig.~\ref{fig5} we show several model bulk phase diagrams (first column) and their corresponding stacking diagrams (second and third columns).  The bulk phase diagram in Fig.~\ref{fig5}a1) consists of a binodal that ends at two critical points. Its corresponding stacking diagram is formed by a sedimentation binodal and two terminal lines, see (b1) and (c1). 

We now replace the upper critical point of the binodal by a phase transition in the pure subsystem of species $2$. This bulk phase diagram is depicted in Fig. \ref{fig5} a2). The binodal tends asymptotically to the value of $\mu_2$ at the transition  ($\lim_{\mu_1\rightarrow-\infty}\mu_2=0.89$). In the stacking diagram, see (b2) and (c2), we observe three types of boundaries: (i) a sedimentation binodal (obtained as the Legendre transform of the bulk binodal), (ii) a terminal line (all paths that cross the critical point), and (iii) an asymptotic terminal line (due to the phase transition of the pure species $2$). The asymptotic terminal line is a vertical line ($s=0$) in the $s$,$a$ representation of the stacking diagram, Fig. b2). It represents all the sedimentation paths that are parallel to the asymptotic behaviour of the binodal in the bulk phase diagram (i.e., paths with a horizontal slope). In the $\alpha$,$r_{\text{min}}$ plane of the stacking diagram, (c2), the asymptotic terminal line is also a vertical line at $\alpha=0$ and $\alpha=\pi$. Here not only the slope but also the direction of paths with horizontal slope in the bulk phase diagram is represented and hence there are two asymptotic terminal lines. 

In panel (a3) of Fig. \ref{fig5} we show a model bulk phase diagram similar to that in (a1), but in which we have replaced the lower critical point by a phase transition of species $1$. As a result the binodal has a vertical asymptote and tends to the value of $\mu_1$ at the transition ($\lim_{\mu_2\rightarrow-\infty}\mu_1=0.89$). The asymptotic terminal line corresponding to this transition is not visible in the $s$,$a$ plane of the stacking diagram, (b3), because it is vertical and hence $s\rightarrow\infty$. In this case the $\alpha$,$r_{\text{min}}$ plane, (c3), is more suitable to represent the stacking diagram. Here the asymptotic terminal line is represented by vertical lines at $\alpha=\pi/2$ and $\alpha=3\pi/2$ (sedimentation paths with vertical slope in the bulk phase diagram).

For completeness we show in (a4) the bulk phase diagram in which the binodal connects both phase transitions, one for each pure species. The stacking diagram, Fig. \eqref{fig5} (b4) and (c4), is composed of a sedimentation binodal and two asymptotic terminal lines. No terminal line is present as there is no bulk critical point.

In all the examples of Fig.~\ref{fig5} the set of possible stacking sequences is the same. However, the topology of the stacking diagram is quite different, as it depends not only on the number of stable phases in bulk, but also on the details of the bulk binodals such as curvature, ending points and asymptotic behaviour. 

\subsection{A realistic phase diagram}\label{realistic}
\begin{figure*}
\epsfig{file=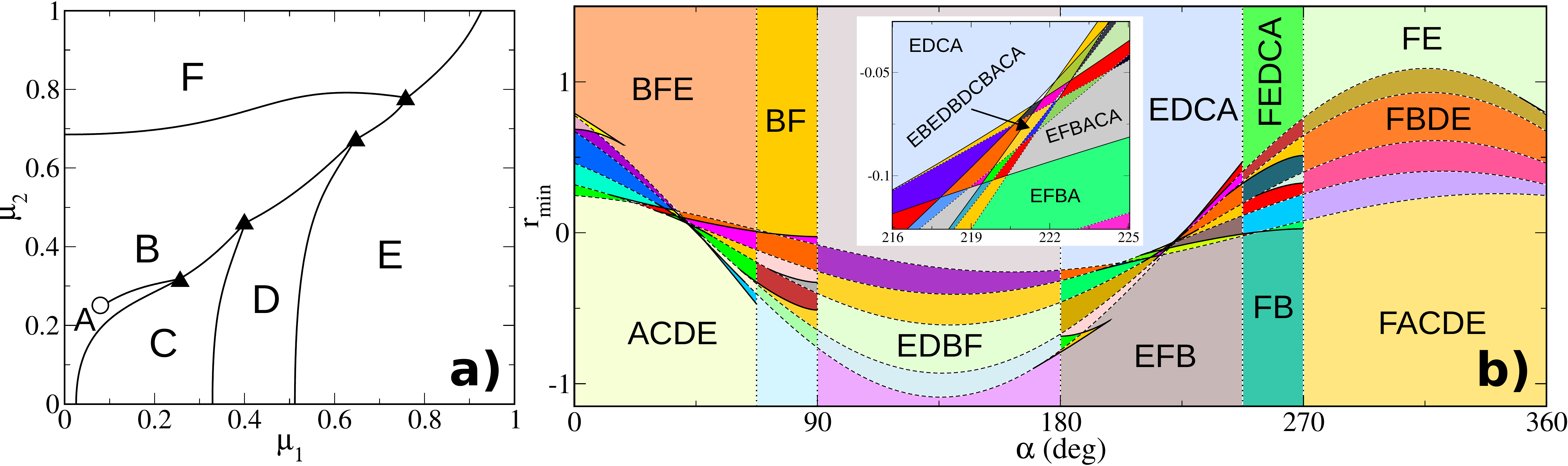,width=1\linewidth,clip=}
\caption{a) Schematic bulk phase diagram of a model binary mixture in the plane of chemical potentials $\mu_1$,$\mu_2$. Chemical potentials are given in arbitrary units. Binodal lines are represented by black-solid lines. The empty circle is a critical point. The black triangles are triple points. (b) Stacking diagram of the mixture in the $\alpha$,$r_{\text{min}}$ plane. Black-solid lines are sedimentation binodals. Dashed lines are terminal lines. Dotted vertical lines are asymptotic terminal lines. The inset is a zoom of a small region of the stacking diagram. Each phase has been colored and some of them are labelled with the stacking sequence. Although some colors are repeated, each region correspond to a distinct stacking sequence.} 
\label{fig6}
\end{figure*}

Real mixtures are by far more complex and typically possess many more than the two stable bulk phases considered above. We hence consider the stacking diagram of a model mixture with six stable bulk phases in order to illustrate the complexity of the stacking diagram that we can expect in real mixtures. The chemical potential representation of the bulk phase diagram is shown in Fig. \ref{fig6}a). The monocomponent system of species $1$ undergoes three first order phase transitions by increasing $\mu_1$: A-C, C-D, and D-E. The species $2$ undergoes only a B-F first order phase transition. At very high chemical potentials there is a large demixing region between phases E and F. At intermediate chemical potentials there is a transition between phases A and B (see the binodal ending at a critical point). Four different triple points occur. A bulk phase diagram like the present one might be representative of, for example, a colloidal mixture of hard spheres and highly anisotropic rods. The sequence of phases for rods (species $1$) by increasing the $\mu_1$ could represent isotropic-nematic-smectic-solid, and for the pure spheres (species $2$) there is a liquid-solid phase transition. A demixing between two isotropic phases, one rich in rods and the other rich in spheres, might correspond to the phase transition between A and B in Fig. \ref{fig6}a).

The corresponding stacking diagram in the $\alpha$,$r_{min}$ plane is depicted in panel (b) of Fig. \ref{fig6}. The boundaries in the stacking diagram are determined by:
\begin{itemize}
\item 18 sedimentation binodals: two sedimentation binodals per each bulk binodal because both directions of the sedimentation paths are considered.
\item 5 terminal lines: one for each triple point ($4$ in total) and one further one due to the critical point.
\item 6 terminal lines: the A-B, C-D, D-E bulk binodals have a vertical asymptote and generate an asymptotic terminal line at $\alpha=\pi/2$ and $\alpha=3\pi/2$. The F-B bulk binodal has a horizontal asymptote and its corresponding asymptotic terminal line is located at $\alpha=0$ and $\alpha=\pi$. We assume that the F-E binodal has an oblique asymptote, and therefore it generates an asymptotic terminal line (located at $\alpha=0.37\pi$ and $\alpha=1.37\pi$).
\end{itemize} 

The stacking diagram is extremely rich, with more than 100 different stacking sequences. The complexity is particularly evident in the region where the slope of the sedimentation path matches approximately the slope of the demixing region in the bulk phase diagram (i.e. a line that connects the triple points in the $\mu_1$,$\mu_2$ representation). This region is represented as an inset in panel (b) of Fig. \ref{fig6}. About 40 different sedimentation phases are possible in this tiny portion of the stacking diagram, such as for example the exotic EBEDBDCBACA sequence with 11 layers. This number is larger than the stable number of phases in bulk because several phases reenter the sequence. Hence, an interesting question arises: what is the maximal number of stacks that can appear in a mixture under gravity? We address this question in the following.

\subsection{Extended Gibbs phase rule for mixtures under gravity}
The Gibbs phase rule determines the maximal number of phases that can coexist in bulk. As discussed above, two non-consecutive phases in a stacking sequence may or may not coexist in bulk. Hence, the number of stacks in a given stacking sequence is not limited by the Gibbs phase rule. Instead, the maximal number $N_{\text{max}}$ of stacks in sedimentation-diffusion equilibrium of a binary mixture is given by
\begin{equation}
N_{\text{max}}=3+2(n_b-1)+n_i,
\end{equation} 
where $n_b$ is the total number of binodals that are present in the bulk phase diagram and $n_i$ is their total number of inflection points. In order to reach $N_{\text{max}}$, a sedimentation path has to cross each binodal $2+n_{b,i}$ times, where $n_{b.i}$ is the number of inflection points of the binodal. In general, $N_{\text{max}}$ is a much larger number compared to the maximal number of possible phases coexisting in bulk ($3$ in an athermal binary mixture). For example, in a mixture with a triple point there are at least three binodals, which sets $N_{\text{max}}\ge7$. Therefore, it might not be surprising to find experimentally colloidal mixtures with sequences that contains $6$ stacks, like the one in~\cite{doi:10.1021/la804023b}.

\subsection{Finite size effects}
Until now we have discussed the stacking diagram considering the sedimentation paths as infinitely long straight lines. We next turn to the case of a mixture in a vessel with finite height, $h$. As a result, any sedimentation path is a line segment (of finite length) in the plane of chemical potentials. The length of the path is proportional to $h$. The main effect of the finite sample is that the stacking diagram becomes even richer in that new stacking sequences appear. These are the formed by removing layer(s) on top or bottom of the stacking sequence corresponding to $h\rightarrow\infty$.

As a concrete example we study a mixture of nonadsorbing polymers and hard core colloidal platelets. We use a perturbation theory developed by Zhang {\it et al.} to study the phase behaviour of the mixture. We refer the reader to Ref. \cite{zhang:9947} for all the details about the theory. The platelets are modeled by cut spheres with diameter $D$ and thickness $L$, that is, spheres of diameter $D$ cut by two parallel planes equidistant from an equatorial plane and separated by a distance $L$. The polymers are treated using the Asakura-Oosawa-Vrij model~\cite{oosawa:1255,vrij}, i.e.\ ideal polymer spheres that cannot overlap with the colloidal platelets. We study the case in which the aspect ratio of the colloidal platelets is $L/D=0.05$ and the size ratio of polymer and colloid is $\sigma_p/D=0.355$, where $\sigma_p=2R_p$ and $R_p$ is the radius of gyration of the polymers. These values are comparable to those in the experimental system analysed in~\cite{PhysRevE.62.5397}. The sedimentation of this mixture has also been previously studied in~\cite{0295-5075-66-1-125} using a different approach.

\begin{figure*}
\epsfig{file=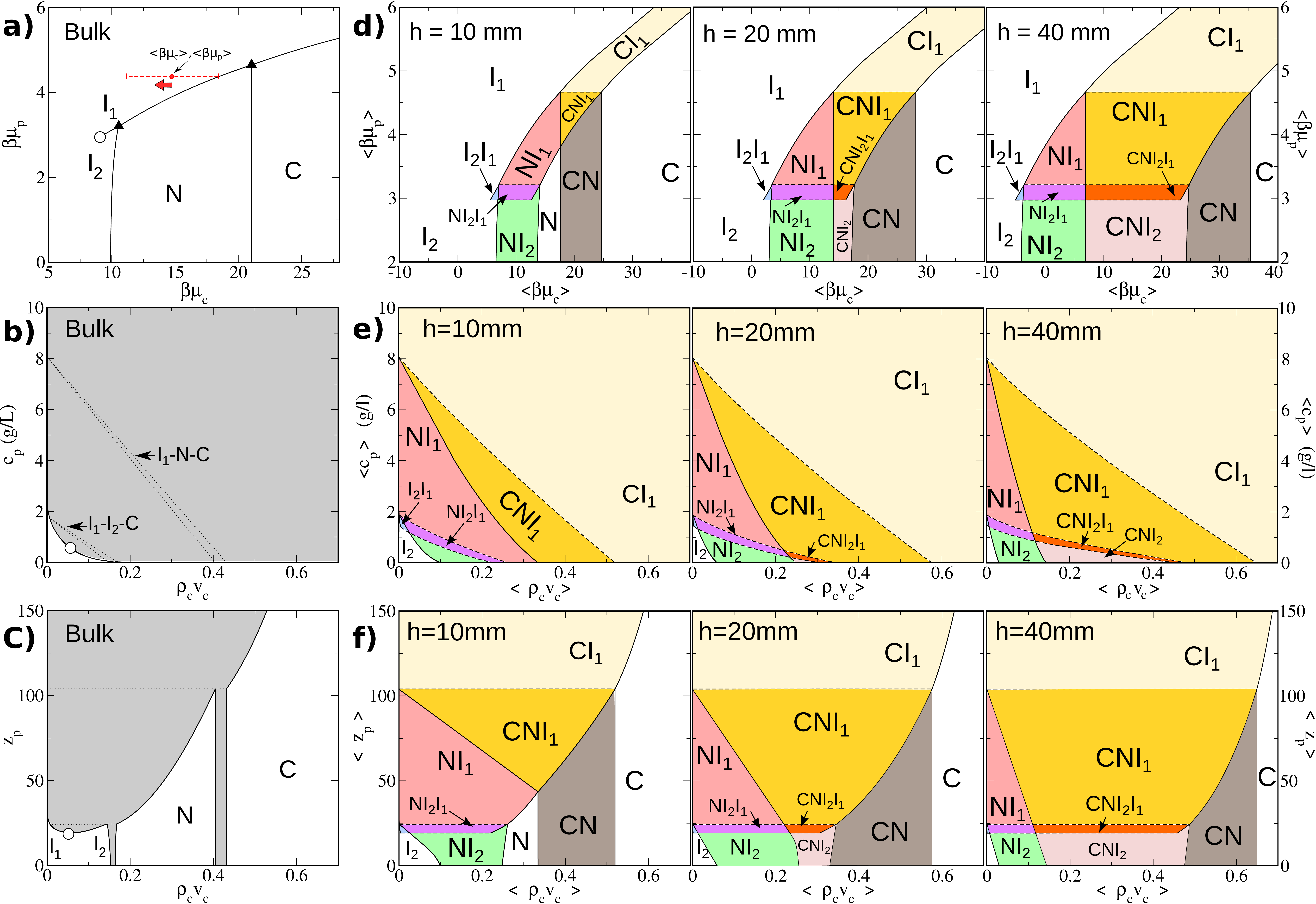,width=1.\linewidth,clip=}
\caption{(left column) Bulk phase diagram of a mixture of nonadsorbing polymers and colloidal platelets in the plane of: (a) chemical potential of colloids $\mu_c$ and polymers $\mu_p$, (b) packing fraction of colloids, $\rho_cv_c$, and concentration of polymers $c_p$ in grams per liter, and (c) packing fraction of colloids $\rho_cv_c$ and fugacity of polymers $z_p$. Empty circles are critical points. Filled triangles (a) and dotted lines (b-c) are triple points. Solid lines are binodals. The grey areas in (b-c) are two-phase regions. The dashed red line in (a) is a sedimentation path of sample with $h=10\text{ mm}$ that corresponds to a boundary between two stacking sequences in the stacking diagram. The coordinates of its middle point are equal to the average chemical potential along the path. (right column) Stacking diagram of the mixture in a vessel with height $h=10$ mm (left), $20$ mm (center), and $40$ mm (right). The different planes are the same as for the bulk phase diagrams (a-c) but considering the average of the quantity of interest along the sedimented sample. E.g., $\langle\rho_cv_c\rangle$ is the average packing fraction of colloids in the vessel. Solid lines are sedimentation binodals and dashed lines are terminal lines. Each region has been coloured and labelled according to the stacking sequence (from bottom to top of the sample).}
\label{fig7}
\end{figure*}

In Fig.~\ref{fig7} we show the bulk phase diagram of the mixture. In (a) we use the plane of chemical potentials, $\mu_c$,$\mu_p$, where $\mu_c$ is the chemical potentials of the colloidal platelets and $\mu_p$ is that of the polymers. In (b) the bulk phase diagram is represented in the plane of polymer concentration, $c_p$, and packing fraction of colloids, $\rho_cv_c$, where $\rho_c$ is the number density of the colloids and $v_c$ the volume of one colloid. To obtain $c_p$ we set the molar mass of the polymers to $4.2\times10^5\text{ g/mol}$, which corresponds to trimethylsiloxy-terminated polydimethylsiloxane  used in~\cite{PhysRevE.62.5397}.  Finally, in (c), we use the plane of fugacity of the polymers, $z_p$, and packing fraction of platelets, $\rho_cv_c$. Here, $z_p=\exp(\beta\mu_p)$ with $\beta=1/k_BT$. The monocomponent system of colloids (i.e., $\mu_p\rightarrow-\infty$, $c_p=0$ and $z_p\rightarrow0$) undergoes two first order phase transitions: isotropic-nematic, I$_2$-N, and nematic-columnar, N-C. The addition of polymers gives rise to new phenomenology: (i) a demixing region occurs at low polymer concentration between two isotropic states I$_1$-I$_2$ bounded by a lower critical point, (ii) two triple points, I$_1$-I$_2$-N, and I$_1$-N-C, and (iii) two demixing regions, I$_1$-N and I$_1$-C, at intermediate and high polymer concentrations, respectively. The I$_1$-C binodal might either end in an isotropic-columnar-crystal triple point or continue for very high chemical potentials. Here we assume the second scenario in order to obtain the stacking diagrams. 

\begin{figure}
\epsfig{file=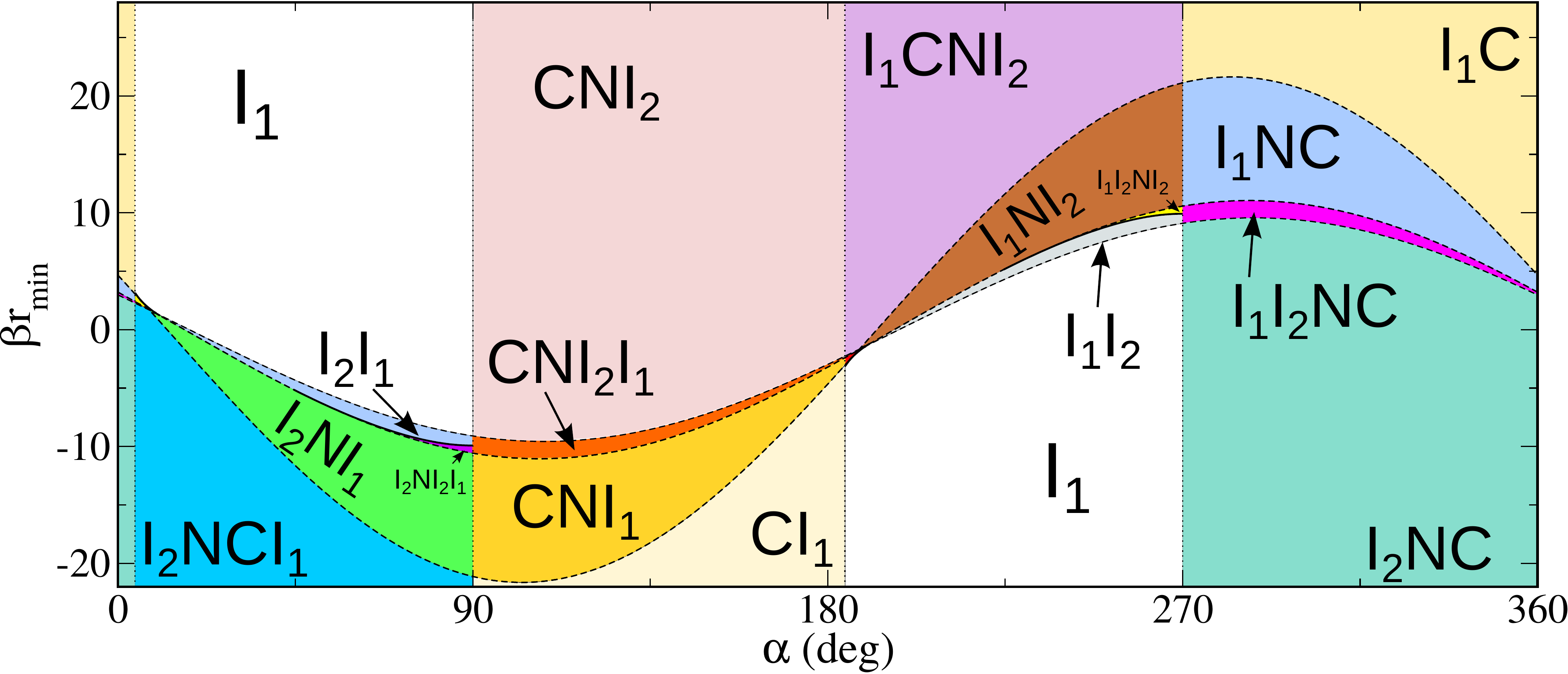,width=1.\linewidth,clip=}
\caption{Stacking diagram of a mixture of nonadsorbing polymers and colloidal platelets in the $r_{\text{min}}$,$\alpha$ plane. Solid lines are sedimentation binodals. Dashed lines are terminal lines. Vertical dotted lines are asymptotic terminal lines. The regions that cover a significant portion of the stacking diagram have been coloured and labelled according to their stacking sequences (from bottom to top of the sample).}
\label{fig8}
\end{figure}

We use the chemical representation of the bulk phase diagram to obtain the stacking diagram in the $\alpha$,$r_{\text{min}}$ plane, see Fig.~\ref{fig8}. For the case of infinite sample height we have identified already 24 different stacking sequences (some of them are not labelled in the figure due to the tiny portion of the stacking diagram in which they are stable). 

As discussed in Sec.~\ref{sedpath}, the stacking diagram for finite sample height requires 5 variables to fully specify each (finite) sedimentation path. In order to simplify this parameter space we fix the slope, the length and the direction of the path. We then vary the location of the path in the plane of chemical potentials by varying the average local chemical potentials of each species $\langle\mu_i\rangle$, with $i={p,c}$. Here $\langle\mu_i\rangle$ is the local chemical potential in the middle of the sample, i.e. at height $z=h/2$. We set the buoyant mass of the colloids to $m_c=2.93\times10^{-6}\text{ g}$, equivalent to a gravitational length $\xi_c=1.41\text{ mm}$ at room temperature ($m_c$ has been calculated for gibbsite platelets dispersed in toluene using the particle dimensions specified in~\cite{PhysRevE.62.5397}). The buoyant mass of the polymers, $m_p$, is negligible compared to that of the colloids and hence we fix the slope of the path to $s=m_p/m_c\rightarrow0$. The local chemical potential of the colloids decreases from bottom to top because $m_c>0$, which fixes the direction of the sedimentation path. We consider three different values of height, $h=10,20$ and $40\text{ mm}$. A sedimentation path that starts or ends at a binodal line forms a boundary in the stacking diagram (see an example in Fig.~\ref{fig7}a) for the case $h=10\text{ mm}$). Such paths form the sedimentation binodals in the stacking diagram. The terminal lines are given by those paths that cross an ending point of a binodal in the chemical potential representation. No asymptotic terminal lines are present because the slope of the path is prescribed and hence does not vary.

In Fig.~\ref{fig7}d) we show the stacking diagram for finite $h$ in the $\langle\mu_c\rangle$,$\langle\mu_p\rangle$ plane. It is interesting to compare the number of possible stacking sequences with the limiting case $h\rightarrow\infty$ (vertical line at $\alpha=\pi$ in Fig. \ref{fig8}). For $h\rightarrow\infty$ there are 4 possible stacking sequences: CI$_1$, CNI$_1$, CNI$_2$I$_1$, and CNI$_2$. However, the diagrams for finite height are much richer. New stacking sequences appear, which are subsequences (i.e. truncated versions) of a corresponding infinite path. The height of the sample is a crucial parameter which modifies the stacking diagram not only quantitatively but also qualitatively. An example is the sequence CNI$_2$ that occurs in the case of $h=20$ and $40\text{ mm}$ but not for $h=10\text{ mm}$, because in the latter case the paths are too short to simultaneously cross both binodals C-N and I-N$_2$. As a result the stacking sequence CNI$_2$ is replaced by N in samples with small $h$.

Once the chemical potential representation of the stacking diagram is known, we can transform it into further representations, such as into the plane of polymer concentration and packing fraction of colloids (Fig.~\ref{fig7}e), or the plane of polymer fugacity and packing fraction of colloids (Fig.~\ref{fig7}f). Here we need to average the quantity of interest along the sedimentation path. For example, to obtain the packing fraction of colloids in the sedimented sample we use
\begin{equation}
\langle\rho_cv_c\rangle=\frac{v_c}{h}\int_0^hdz\rho_c(z),
\end{equation}
where $\rho_c(z)$ is the density profile of the colloids as a function of the vertical coordinate. The density profile can be obtained using the equation of state $\rho_c(\mu_c)$ and the values of $\mu_c$ along the sedimentation path. 

A quantitative comparison with the experimental observations in~\cite{PhysRevE.62.5397} cannot be performed, because of variation in the height of the experimental samples, possibly due to solvent evaporation. However, the same stacking sequences are present in both experiments and theory. Our theory predicts also the same stacking sequences than those in~\cite{0295-5075-66-1-125} in which the sedimentation of the same mixture is considered by analyzing osmotic equilibrium conditions in the mixture.  

\subsection{Topologies of the bulk phase diagrams}
\begin{figure}
\epsfig{file=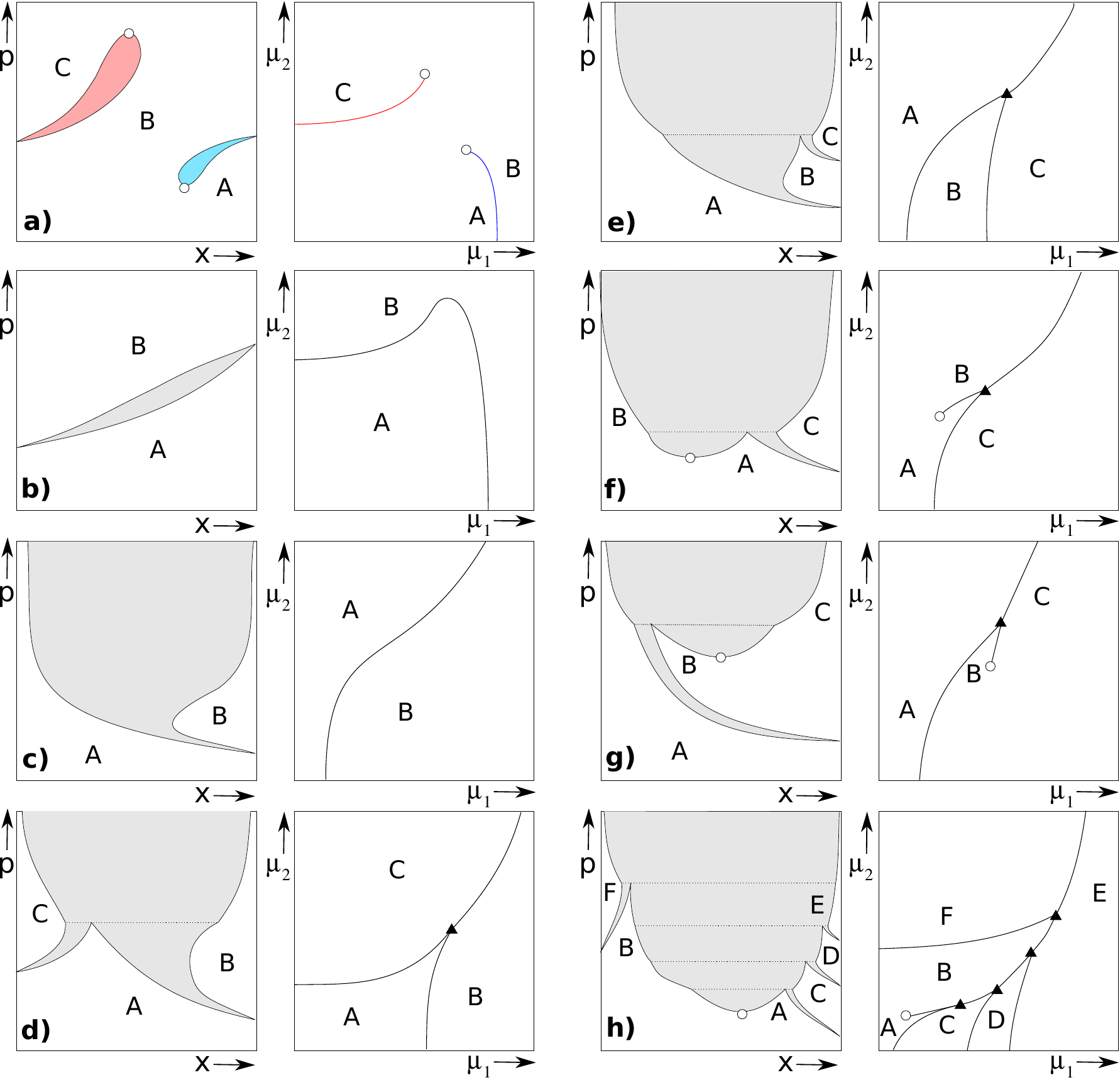,width=1\linewidth,clip=}
\caption{(a)-(h) Schematic bulk phase diagrams in the composition $x$, pressure $p$ plane (left) and their corresponding bulk phase diagrams in the plane of chemical potentials $\mu_1$,$\mu_2$ (right). $x$ is the molar fraction of species $1$. The shadow areas represent two-phase regions. Solid lines are bulk binodals. Empty circles denote critical points. Filled triangles and horizontal dotted lines indicate triple points in the $\mu_1$,$\mu_2$ and $x$,$p$ representations, respectively. }
\label{fig9}
\end{figure}

As the sedimentation paths are described by straight lines in the chemical potential representation of the bulk phase diagram, this representation is particularly suited for the analysis of sedimentation-diffusion equilibrium. However, in the literature, the chemical potential is certainly not the most commonly used thermodynamic variable to represent the bulk phase diagram. Other representations of the bulk phase diagram, such as via the composition, $x$, pressure, $p$, plane are by far more popular. In order to connect both representations, $x$,$p$ and $\mu_1$,$\mu_2$, we plot in Fig.~\ref{fig9} a catalog of schematic bulk phase diagrams in both variable sets. Here we define $x$ as the molar fraction of the species $1$ and, in colloidal mixtures, $p$ is the osmotic pressure. The topological correspondence between both planes is accurate (i.e., number of binodals, ending points, asymptotic behaviour of the binodals...), but certain details, such as the precise curvature of the binodals, depend on the system under consideration and are, therefore, approximate.  

The simplest phase diagram in the $x$,$p$ plane (not shown) is a closed loop of immiscibility, which corresponds to a binodal that ends at two critical points in the plane of chemical potentials, similar to the case presented above. Closed loops of immiscibility in mixtures are infrequent, but they have been predicted in e.g., mixtures of hard squares~\cite{PhysRevE.76.031704} and patchy colloids~\cite{C0SM01493A}.

In Fig.~\ref{fig9}a) we show results for a mixture in which both pure components undergo a phase transition. Both transitions extend into the mixed system within a finite range of composition and then end at critical points. In (b) each pure component again undergoes a transition, but these transitions are connected in the mixture. The stacking diagram of a mixture with a similar phase diagram (mixture of hard platelets with different radius) was analysed in~\cite{stackingdiagram}. In (c) only species $1$ has a transition that extends as a large demixing region at high pressure/chemical potentials. In (d) species $1$ undergoes a transition between phases A and B.  Species $2$ has a A-C transition. Both transitions collapse in a A-B-C triple point. At pressure/chemical potentials above the triple point there is a demixing region between phases B and C. A variant of this phase diagram is depicted in (e). Here the species $1$ undergoes both transitions A-B and B-C. In (f) and (g) we introduce a new element, a demixing region bounded by a lower critical point. Finally in (h) we show a complex phase diagram with 6 stable phases in bulk. The stacking diagram of a similar system has been studied in Sec.~\ref{realistic}. This catalog demonstrates that despite its slight unfamiliarity, the plane of chemical potentials is a suitable means to represent common bulk phase phenomenology.

\section{Discussion}
We have developed a theory of the phenomenology of stacking sequences that occur in sedimentation-diffusion equilibrium of colloidal mixtures, based on a local density approximation and the concept of the sedimentation path, which is a straight line in the plane of chemical potentials. An interface in the sample corresponds to a crossing between the path and a binodal in the chemical potential representation of the bulk phase diagram. Hence a direct relation is established between the sedimentation path and the observed stacking sequence in the vessel. We have grouped all possible stacking sequences in a well-defined mathematical object, the stacking diagram. Three elements form the boundaries between different stacking sequences in the stacking diagram: (i) sedimentation binodals, (ii) terminal lines, and (iii) asymptotic terminal lines. The bulk phase diagram and the stacking diagram are linked via a mathematical correspondence based on the Legendre transform. 

As the gravitational length is in general much larger than all relevant correlation lengths in the system, we expect our "local density approximation" approach to be very accurate in general. Actually, it has been shown to be an excellent approximation in mixtures of colloidal platelets and spheres~\cite{floating}. Nevertheless, the LDA treatment might be inaccurate in presence of long range pair interactions, and will fail if the largest correlation length in the system is comparable to the shortest gravitational length of the components of the mixture. In addition, effects such as wetting at the upper and lower interfaces of the vessel, and interactions between two consecutive interfaces of the stacking sequence, go beyond the local density approximation. These nonlocal effects will enrich and modify the stacking diagram by e.g., the the occurrence of new boundaries due to e.g. prewetting transitions.  

As we have shown, finite size effects enrich also the stacking diagram. We have considered an example of stacking diagram for finite heights in a mixture of polymers and colloids. In this case the sedimentation path is a line segment and as  a result of its finite length new stacking sequences appear (compared to the case of an infinite sample). The height of the sample determines the length of the path. Hence, the container height is a key variable that should be carefully controlled in experiments. 

Another key parameter is the ratio between the buoyant masses of both species (slope of the sedimentation path). This ratio might be varied by changing the solvent of the mixture. Although we have focused on colloidal mixtures, the theory is also valid for molecular mixtures. The gravitational length in molecular systems is several orders of magnitude higher than that in colloidal systems. Then, our theory might be relevant to understand the structure of geological deposits~\cite{Debenedetti1988,EspositoEtAl2000}.

In multicomponent systems the sedimentation paths remain straight lines in the phase space of chemical potentials. Hence, the theory can be extended to study the stacking diagrams in systems with more than two components. The sedimentation of multicomponent mixtures has been recently investigated in~\cite{multi,0295-5075-89-3-38006}.

  We have considered here the case where gravity is the
  only external potential. However, one can image mixtures with
  additional external potentials~\cite{0953-8984-20-40-404201} or
  pseudo-external potentials, i.e., a potential self generated by the
  components of the mixture. Examples are colloids in presence of
  external electric fields ~\cite{0953-8984-15-1-302,C4SM00957F},
  charged colloids~\cite{racsa2004evidence,leunissen2005ionic}, and
  dense polymer solutions~\cite{wittmer2004long}. If the additional
  external potential is linear in $z$, then the theory remains valid
  with a redefinition of the intersection $a$ and the slope $s$ of the
  sedimentation path. If the additional external potential is not
  linear in $z$, then the sedimentation path is, in general, no longer
  a straight line in the plane of chemical potentials. Our concept of
  identifying the crossings between such a curved sedimentation path
  and the characteristics of the bulk phase diagram remain valid,
  however the (geometric) generalization from straight to curved paths
  needs to be performed.

The study of sedimentation-diffusion equilibrium using computer simulations is a time consuming task because gravity affects the system at scales much larger than the particle length. Our approach can be relevant in computer simulations because only the bulk phase diagram of the mixture needs to be simulated in order to predict the sedimentation-diffusion equilibrium regime.

\end{document}